\documentclass[12pt]{article}

\usepackage{graphicx}        
\usepackage{epsfig} 
\usepackage{psfrag}
\usepackage{rotate}
\usepackage{amssymb}
\usepackage{amsfonts}
\usepackage{latexsym}
\usepackage{amsmath}
\usepackage{cite}
  
\newcommand{\be}{\begin{equation}}
\newcommand{\bea}{\begin{eqnarray}}
\newcommand{\bdm}{\begin{displaymath}}
\newcommand{\ee}{\end{equation}}  
\newcommand{\eea}{\end{eqnarray}}
\newcommand{\edm}{\end{displaymath}}

\begin{document}

\pagestyle{empty} 

\begin{flushright}
hep-lat/0503021
\end{flushright}
 
\vspace{0.5cm}
\begin{center}
\boldmath{\Large \bf Casimir Scaling in \\
$SU(2)$ Lattice Gauge Theory}
\unboldmath\\
\vspace{1cm}
{\large
Carlo~Piccioni}
 
\vspace{0.5cm}
{\it Department of Physics\\
 New York University\\
4 Washington Place\\
New York NY 10003, USA} 
\end{center}
\bigskip
 
\begin{center}
\bf Abstract
\end{center}

{By employing  the multilevel algorithm in numerical Monte Carlo simulations, we evaluate  the static potential in four dimensional $SU(2)$ lattice gauge theory with no dynamical fermions, for static sources in the $j=\frac{1}{2},1,\frac{3}{2}$ representations. We find data supporting the Casimir scaling hypothesis. With the same technique we show that the ratio of the logarithm of Wilson loops in the $j=2$ and $j=\frac{1}{2}$ representations, as a function of the area of the loops, also satisfies Casimir scaling.}
 
\vfill
\hrule width 5.cm
\vskip 2.mm
{\small \noindent e-mail: carlo.piccioni@physics.nyu.edu}

\newpage

\setcounter{page}{1}
\pagestyle{plain}   

\section{Introduction}

Numerical evaluations of Wilson loops are of key importance in the lattice approach to the confinement problem, because of the relation between Wilson loops and the energy between static sources. Wilson loops become increasingly small as the  separation of the sources increases. As a result they are hard to measure, because statistical fluctuations cover their signal. When studying confinement this is clearly a problem, because we are interested in the  large distance properties of the sources. For this reason it is common  to find examples in the literature where new algorithms are introduced, in order to either improve the signal, see for example the smearing \cite{Albanese:1987ds}, variational \cite{Michael:1985ne,Luscher:1990ck} and blocking \cite{Teper:1987wt} algorithms, or reduce the statistical fluctuations, see for example the multihit \cite{Parisi:1983hm} and multilevel \cite{Luscher:2001up} algorithms. In particular the multilevel algorithm has been shown to be a very powerful error reduction technique to evaluate small magnitude Wilson loops \cite{Luscher:2001up,Peardon:2002ye,Majumdar:2002mr,Meyer:2002cd}. When considering Casimir scaling  in lattice gauge theory\cite{Ambjorn:1984mb,DelDebbio:1995gc}, the signal is much weaker because higher dimensional  representation Wilson loops should be obtained by exponentiation of fundamental loops with the ratio of the corresponding Casimir operators. The Casimir scaling hypothesis precisely states that the ratio between the static potential in some representation of the gauge group and the static potential in the fundamental representation at the same distance, is equal to the ratio of the corresponding Casimir operators. This should be true from the onset of confinement to the onset of color screening. Using the language of the $k$-string studies performed in \cite{Lucini:2001nv,Armoni:2003nz,Gliozzi:2005dv}, at asymptotically large distances string tensions should be dependent on the $N$-ality $k$ of the representation of the sources rather than the representation itself, and as consequence a string is expected to decay into the corresponding stable $k$-string. There is considerable evidence of Casimir scaling in the non-perturbative regime from Monte Carlo calculations. Studies of the $SU(3)$ case can be found for example in \cite{Deldar:1999vi,Bali:2000un}, while the four and three dimensional $SU(2)$ cases are discussed in \cite{Ambjorn:1984mb,Michael:1985ne} and \cite{DelDebbio:1995gc,Poulis:1995nn,Kratochvila:2003zj} respectively. Nevertheless there is no clear theoretical explanation, in fact Casimir scaling is known to be true in lattice and continuum four dimensional Yang-Mills theories only perturbatively. One may explain Casimir scaling in four dimensions by dimensional reduction arguments \cite{Ambjorn:1984mb}, remembering that it is true exactly in two dimensions. However, as pointed out  in \cite{Brzoska:2004pi}, in two dimensions there is no color screening and therefore the asymptotic behavior of Wilson loops at large distances is different. Other models have been proposed, based on arguments on  vortex theory \cite{Faber:1997rp} and  diffusion of Wilson loops \cite{Brzoska:2004pi}. Evidence of exponentiation of Wilson loops with Casimir operators can also be found in \cite{Zwanziger:1998ez}. 

In this work we apply the multilevel algorithm to the study of Casimir scaling in four dimensional $SU(2)$ lattice gauge theory. In this case, as in $SU(3)$, the only non trivial string tension at large distances is the fundamental, because we can only have $k=0,1$ corresponding to the trivial and the fundamental representations respectively. Nevertheless it is still interesting to study the static potential of sources in higher dimensional representations at intermediate distances, where Casimir scaling is conjectured. For $SU(2)$ the ratio of the Casimir operators increases very rapidly with increasing dimension of the representation, and the determination of static potentials is rather difficult. We test the Casimir scaling hypothesis for the $j=\frac{1}{2},1,\frac{3}{2},2$ representations. In section~2 we introduce static potentials and Casimir scaling, in section~3 we explain the technique and show the results, in section~4 we conclude with some remarks.

\section{Casimir scaling of static potentials}

The static potential is the energy $E_0$ of the ground state of a pair of heavy static sources. In the continuum formulation one can show that its relation to the Euclidean expectation value of Wilson loops is given by 
\be
W(R,T)=\langle \Psi(R)|e^{-HT}|\Psi(R)\rangle=\sum_{n=0}^{\infty}|\langle \Psi(R)|\Phi^n(R)\rangle|^2 e^{-E_n(R)T},
\ee
where $|\Psi(R)\rangle$ is the state describing  the pair at distance $R$ and  $H$ is the Hamiltonian of the system with orthonormal eigenstates $|\Phi^n(R)\rangle$, and eigenvalues $E_n(R)$. The state $|\Psi(R)\rangle$ is not an eigenstate of the Hamiltonian. We can extract information on the ground state energy $E_0(R)=V(R)$ by taking the limit $T\to \infty$ where the sum is dominated by the first term, obtaining
\be
W(R,T)=C_0(R) e^{-V(R)T},
\ee
where $C_0(R)=|\langle \Psi(R)|\Phi^0(R)\rangle|^2 $ is the overlap of $|\Psi(R)\rangle$ with the ground state $|\Phi^0(R)\rangle$. From this equation we obtain the  static potential 
\be
V(R)=-\lim_{T\to \infty} \frac{\ln W(R,T)}{T}.
\ee
On a lattice of spacing $a$ Eqs.(2,3) hold  with $R=\hat{R}a$, $T=\hat{T}a$. 

We work with the $SU(2)$ lattice Yang-Mills theory in the quenched approximation, described by the lattice action
\be
S(U)=\beta\sum_{\Box}(1-\frac{1}{2}TrU_{\Box})
\ee
where $U_{\Box}$ is the fundamental plaquette, $\beta=\frac{4}{g_0^2}$ and $g_0$ is the lattice  coupling constant. We consider the Euclidean expectation  values $W(R,T)$ of Wilson loops $W_C(U)=\frac{1}{2}Tr(\prod_{l\in C}U_l)$ around rectangular loops $C$ of spatial extension $R$ and temporal extension $T$. It is possible to consider sources in the higher dimensional representations  $D^j$, $j=0,\frac{1}{2},1,\frac{3}{2},2,...$, of $SU(2)$ and the corresponding Wilson loops  obtained replacing each link of the loop  with its representation. From the definition of representation and the character of $SU(2)$ representations we have
\be
W_C^j(U)=\frac{1}{2j+1}Tr\left[D^j(U_{\frac{1}{2}})\right]=\frac{1}{2j+1}\frac{\sin\left[(2j+1)\frac{\phi}{2}\right]}{\sin(\frac{\phi}{2})},
\ee
where $U_{\frac{1}{2}}=\prod_{l\in C}U_l$ and $\phi$ is obtained expanding the element $U_{\frac{1}{2}}\in SU(2)$ in terms of the Pauli matrices as $U_{\frac{1}{2}}=\cos(\frac{\phi}{2})I_2+i\sin(\frac{\phi}{2})\hat{n}\cdot \vec{\sigma}$. This formula allows us to calculate $W_C^j(U)$ once we find the parameter $\phi$ of  $U_{\frac{1}{2}}$. From Eqs.~(2,3), written in terms of $W_j(R,T)=\langle W_C^j(U)\rangle$, we can evaluate the static potential $V_j(R)$ between  static sources in the representation $j$.

According to the Casimir scaling hypothesis, \cite{Ambjorn:1984mb,DelDebbio:1995gc}, the ratio between the static potential in some representation and the static potential in the fundamental representation at the same distance, is equal to the ratio of the corresponding Casimir operators
\be
\frac{V_j(R)}{V_{\frac{1}{2}}(R)}=\frac{C_j}{C_{\frac{1}{2}}}=\frac{4}{3}j(j+1),
\ee
where $C_j=j(j+1)$ is the Casimir operator of the representation $j$. If  the static potential is linearly rising with the distance and the Casimir scaling hypothesis holds true, we have $V_j(R)= \sigma_j R$, where $\sigma_j$ is the string tension in the representation $j$, and therefore $\sigma_j=\frac{4}{3}j(j+1)\sigma_{\frac{1}{2}}$. From Eq.~(3) we have  for Wilson loops in the representation $j$, $\ln W_j(R,T)=\frac{4}{3}j(j+1)\ln W_{\frac{1}{2}}(R,T)$ at large $T$, from which we find
\be
W_j (R,T)=\left[W_{\frac{1}{2}}(R,T)\right]^{\frac{4}{3}j(j+1)}.
\ee
We see immediately that the  Wilson loops decrease dramatically as $j$ increases. For example if a Wilson loop in the fundamental representation is of the order of $10^{-2}$, then in the $j=1,\frac{3}{2},2$ representations should be of the order of $10^{-6},10^{-10},10^{-16}$ respectively. To measure Wilson loops of these small magnitudes, it is necessary to introduce algorithms that either enhance their signal or decrease the statistical fluctuations. This is the subject of the following section. 

\section{Technique and results}

On the lattice we can write Eq.(2)  for the representation $j$ in the form  $W_j(\hat{R},\hat{T})=C_0^j(\hat{R})e^{-\hat{V_j}(\hat{R})\hat{T}}$, where we have introduced the static potential  in lattice units $\hat{V_j}(\hat{R})=V_j(\hat{R}a)a$. Then the static potential in lattice units is given by
\be
\hat{V}_j(\hat{R})=-\lim_{\hat{T}\to \infty}\ln\left[\frac{W_j(\hat{R},\hat{T}+1)}{W_j(\hat{R},\hat{T})}\right]. 
\ee
In practice for fixed $\hat{R}$ we valuate the quantities $\hat{V_j}(\hat{R},\hat{T})=-\ln\left[\frac{W_j(\hat{R},\hat{T}+1)}{W_j(\hat{R},\hat{T})}\right]$ and extrapolate $\hat{V}_j(\hat{R})$ when a plateau value in $\hat{T}$ is reached. We use lattices of size $16^4$ and $12^3\times 16$ for $j=\frac{1}{2}$ and $j=1$ respectively, and lattices of size $8^3\times 12$ for $j=\frac{3}{2}$ and $j=2$, at $\beta=2.5$. We update the gauge fields by use of the heat bath algorithm introduced in \cite{Creutz:1980zw}. To enhance the projection on the ground state we use spatial smeared links \cite{Albanese:1987ds} in the Wilson loops, obtained  by the following prescription
\be
U_i(n)\to \frac{U_i(n)+\epsilon S_i}{\sqrt{\det(U_i(n)+\epsilon S_i)}},
\ee
where 
\be
S_i=\sum_{j\ne i}[U_j(n)U_{i}(n+\hat{j})U_j^{\dagger}(n+\hat{i})+U_j^{\dagger}(n-\hat{j})U_{i}(n-\hat{j})U_j(n-\hat{j}+\hat{i})],
\ee
and $i,j=1,2,3$. The procedure is iterated a number $n_s$ of times. We find that the choices of the smearing parameters $\epsilon=0.1, n_s=37$ for $j=\frac{1}{2}$, $\epsilon=0.3, n_s=13$ for $j=1$ and $\epsilon=0.5, n_s=31$ for $j=\frac{3}{2}$  work well. For $j=2$ it is difficult to determine the smearing parameters, in fact the signal is very weak and the dependence of the parameters on the spatial extent of the loops is more sensitive. On the temporal links of the loops, we apply the multilevel algorithm as an error reduction technique. This algorithm was introduced in \cite{Luscher:2001up} as generalization of the multihit algorithm \cite{Parisi:1983hm}, where each link is substituted by its expectation value, obtained keeping the neighboring links fixed. We tried to apply the multihit algorithm with no satisfactory results. 
\begin{figure}[t!]
\begin{center}
\includegraphics{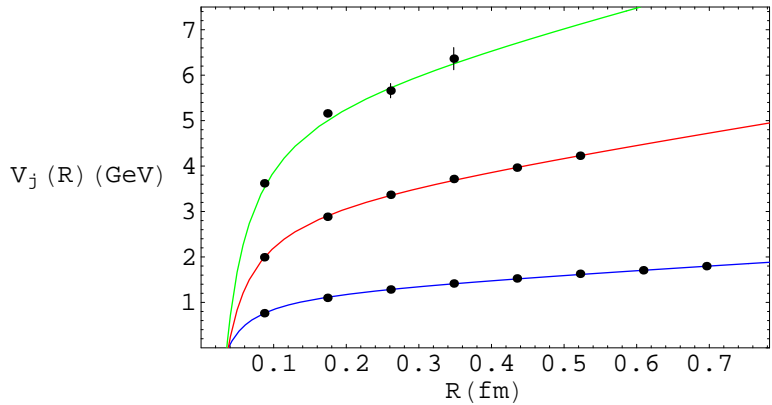} 
\caption{(color online). \sl The physical static potential for the $j=\frac{1}{2},1,\frac{3}{2}$ representations. The dots  are the numerical values obtained, and the solid lines are the potential fits.}
\end{center}
\end{figure}
The multilevel algorithm is employed with a two level scheme according  to which, following the notation and terminology of \cite{Luscher:2001up}, the Euclidean expectation values $W(\hat{R},\hat{T})=\langle W_C(U)\rangle $ are given by
\be
W(\hat{R},\hat{T})=\langle{\mathbf L}(0)_{\alpha \gamma}\{[[{\mathbf T}(0)][{\mathbf T}(1)]]...[[{\mathbf T}(\hat{T}-2)][{\mathbf T}(\hat{T}-1)]]\}_{\alpha \beta \gamma \delta}{\mathbf L}(\hat{T})_{\beta \delta}^*\rangle,
\ee
where ${\mathbf L}$ denotes the  spatial transporters, ${\mathbf T}$ denotes the two-link operators and $[.]$ denotes the time-slices expectation values, obtained keeping fixed the links on the boundary of the time-slices. With this scheme only even temporal extension Wilson loops can be calculated. We evaluate the $[{\mathbf T}]$ averages from $7$ updates of the time-slices of width $1$, and the $[[{\mathbf T}][{\mathbf T}]]$ averages  from  $9$ updates of the time-slices of width $2$. Note that since the Wilson loops are obtained with the expectation values of the two-link operators $\mathbf T$, we cannot make use of Eq.~(5) to evaluate the higher dimensional representation traces, and we must consider the matrix form of the links in the representation $j$. In figure~1 we plot the results obtained for the physical static potentials for $j=\frac{1}{2},1,\frac{3}{2}$. We obtain the following least square fits for the potentials in lattice units
\be
\hat{V}_j(\hat{R})=\hat{V}_0^j-\frac{\hat{e}_j}{\hat{R}}+\hat{\sigma}_j \hat{R},\ee
with the following values for the parameters
\begin{center}
\begin{tabular}{ccc}
$\hat{V}_0^{\frac{1}{2}}=0.54(1)$&$\hat{e}_{\frac{1}{2}}=0.24(1)$&$\hat{\sigma}_{\frac{1}{2}}=0.035(1)$
\\
$\hat{V}_0^1=1.40(3)$&$\hat{e}_1=0.62(3)$&$\hat{\sigma}_1=0.094(5)$
\\
$\hat{V}_0^{\frac{3}{2}}=2.4(4)$&$\hat{e}_{\frac{3}{2}}=1.0(3)$&$\hat{\sigma}_{\frac{3}{2}}=0.15(8)$.
\end{tabular}
\end{center}
\begin{figure}[t!]
\begin{center}
\begin{tabular}{cc}
\includegraphics[angle=0,width=0.45\textwidth]{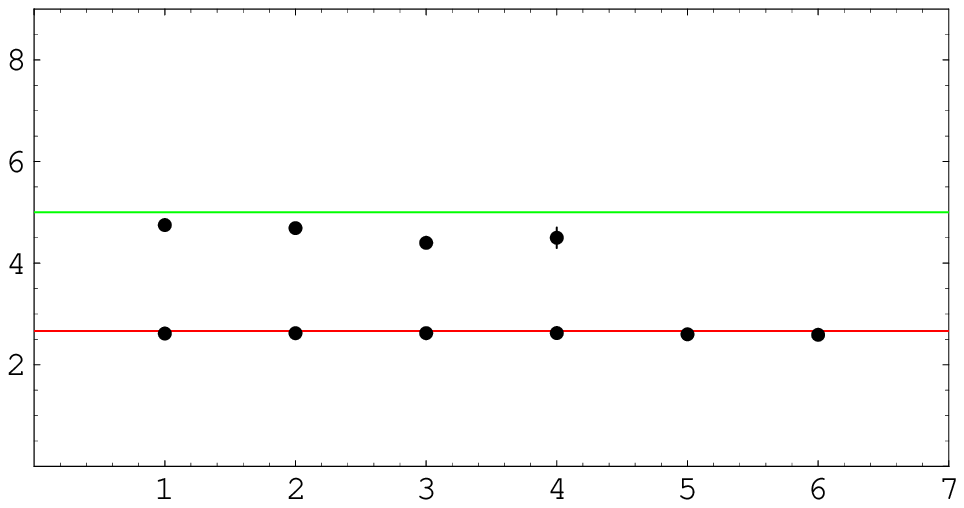} & \includegraphics[angle=0,width=0.45\textwidth]{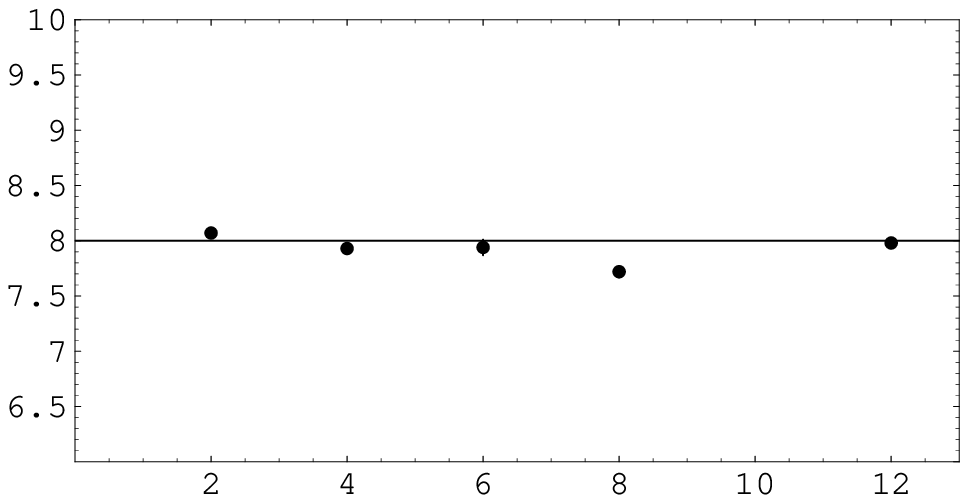} 
\end{tabular}
\caption{(color online). \sl Left: ratio of the static potentials for $j=1$ and $j=\frac{3}{2}$, against the ratio of the corresponding Casimir operators. Right: ratio of the logarithm of Wilson loops for $j=2$, against $\frac{C_2}{C_{\frac{1}{2}}}=8$.}
\end{center}
\end{figure}
The lattice size $a=0.087(1)fm$, or $a=0.440(6)GeV^{-1}$, is extrapolated from the relation  $\sigma_{\frac{1}{2}}=\frac{\hat{\sigma}_{\frac{1}{2}}}{a^2}$, giving to the string tension the physical value $\sqrt{\sigma_{\frac{1}{2}}}=425MeV$. In figure~2 we plot the ratios $\frac{\hat{V}_j(\hat{R})}{\hat{V}_{\frac{1}{2}}(\hat{R})}$ of the static potentials in lattice units, against the ratios of the Casimir operators $\frac{C_1}{C_{\frac{1}{2}}}=\frac{8}{3}$ for $j=1$, and $\frac{C_{\frac{3}{2}}}{C_{\frac{1}{2}}}=5$ for $j=\frac{3}{2}$. For  $j=2$ we plot the ratios $\frac{\ln W_2(\hat{A})}{\ln W_{\frac{1}{2}}(\hat{A})}$ as a function of the area, against $\frac{C_2}{C_{\frac{1}{2}}}=8$. In this case evaluation of the static potential is difficult since the signal is very weak and mostly lost in statistical fluctuations. Exponentiating the fundamental potential rescaled with ratio of the Casimir operators, we find that Wilson loops in the $j=2$ representation should get as small as $\approx 10^{-13}$.

\section{Conclusions and outlook}

The data obtained support the Casimir scaling hypothesis in four dimensional $SU(2)$ lattice gauge theory. By employing  the multilevel algorithm we are able to extract the static potential from Wilson loops, for the $j=\frac{1}{2},1,\frac{3}{2}$ representations. In these cases we test directly the Casimir scaling hypothesis, by considering the ratio of the lattice potentials with the fundamental. In the $j=2$ representation case we obtain an indirect test of Casimir scaling by considering the ratio of the logarithms of Wilson loops as a function of the area. Analysis of the data highlights the multilevel algorithm as a powerful technique to evaluate small magnitude Wilson loops. Although the string picture would be more appropriate to a larger length scale than the one studied in this work, below $0.7fm$, we could not consider larger lattices due to the limited computational availability.

As observed in the introduction, Casimir scaling is expected to be true up to a certain distance between the sources after which the gluonic string holding the pair together should break or be screened to the fundamental, for integer or half integer $j$ respectively. Adjoint string breaking in four dimensional $SU(2)$ lattice gauge theory has been observed in \cite{Michael:1985ne,deForcrand:1999kr}. In \cite{deForcrand:1999kr} the mass of the resulting gluelump is measured and the string breaking scale is set to $R_B=1.25fm$. A rough estimate of the decay scale for the $j=\frac{3}{2}$ representation is set at $\frac{5}{4}$ of the adjoint in \cite{Gliozzi:2005dv}. Since our interquarks separations are much below these decay scales, we cannot test color screening and the $N$-ality dependent string tension scenario. This would certainly be possible using larger size lattices, resulting in new interesting numerical data that, although trivially expected on theoretical grounds, are not yet available in the literature.

\subsection*{Acknowledgments}

We would like to thank Daniel Zwanziger for discussions. Most of the calculations were performed with one Pentium IV $2.66 GHz, 500 MB$ processor. In the last two months of the project we had access to a cluster of Pentium III $1.26 GHz, \sim 1GB$ processors, for which we are grateful to Alan Sokal.

\end{document}